# Investigation of wetting layers in InAs/GaAs self-assembled nanostructures with reflectance difference spectroscopy

H. Y. Zhang, Y. H. Chen,* Z. G. Wang

Key Laboratory of Semiconductor Materials Science,

Institute of Semiconductors, Chinese Academy of Sciences,

P.O. Box 912, Beijing 100083, People's Republic of China

1  **Introduction**
2  **The wetting layer characterization methods**
    2.1  **Direct imaging method**
    2.2  **Spectroscopy method**
    2.3  **Reflectance difference spectroscopy**
3  **Wetting layer evolution in SK mode grown QDs**
    3.1  **A general description of the wetting layer evolution in SK mode**
    3.2  **Segregation effect**
    3.3  **RDS in determining the critical thickness**
    3.4  **Desorption of indium atoms**
    3.5  **Effect of growth interruption**
4  **Wetting layer in Droplet epitaxy mode grown nanostructures**
    4.1  **Discontinuous wetting layers in droplet epitaxy mode grown nanostructures**
    4.2  **Wetting layer evolution in droplet epitaxy mode grown nanostructures**
5  **Summary**

**Abstract**

As a system with both profound physics and promising application potentials, the strain-induced self-assembled semiconductor nanostructures have been investigated for decades of years. The optical and electrical properties of this system are mainly determined by the nanostructures, but can be greatly affected by the generally existent two dimensional structures from which the nanostructures evolve, the so called wetting layer (WL). The WL configurations can be varied with different growth conditions, and further influence the nanostructure morphology and properties. This is a reviewing article introducing some recent progresses in investigating the evolution of WLs in InAs/GaAs system with reflectance difference spectroscopy. Different kinds of WL evolution processes in Stranski–Krastanov growth mode are introduced. The segregation and desorption of indium atoms and the effect of growth interruption are mentioned. The existence and the evolution process of wetting layers in droplet epitaxy method are also discussed.

* Electronic address: yhchen@red.semi.ac.cn

## 1 Introduction

Semiconductor nanostructures, such as quantum dots (QDs), quantum wires (QWRs) and quantum rings (QRs), have attracted great attentions duo to their interesting physical properties and promising application potentials in various kind of optoelectronic devices, such as laser diode, detector, single photon source and entangled photon source, et al.[1-4] Self-assembly on the bulk semiconductor surface is the most common way to obtain the semiconductor nanostructures.

If a certain degree of lattice mismatch exists between the substrate and the epitaxial film, some dislocation-free islands may generate to relax the strain and lead to a more stable system with three-dimensional (3D) QDs on top of a two-dimensional (2D) wetting layer (WL), which is the so-called Stranski–Krastanov (SK) growth mode. SK mode is the most conventional way for the self-assembly of QDs with nonconventional luminescence properties. According to the general description of SK growth mode, there is a critical thickness of the WL as a turning point of the 2D to 3D morphology transition. Above the critical thickness, QDs with a typical areal density of $10^{10}cm^{-2}$ form and the WL tends to be stable. The optical properties of this system are mainly determined by the energy states of the QDs, but can be greatly affected by the wetting layer (WL). For examples, the WLs can serve as channels for the carriers to exchange and redistribute between QDs, which greatly influences the emission properties of the QD ensemble. [5-7] By controlling the WL configurations, the optical properties of QDs can be tuned without changing the QD morphology and composition. [8] The nucleation and evolution of QDs can be modulated by the WL morphology. [9] As to the QD based laser, the gain-current characteristic and modulation response can be greatly influenced by the WL states.[10, 11] It is important to know the structures and electronic states of WLs to understand the properties of those self-assembled nanostructures and their devices.

For the SK growth mode, the detailed WL evolution processes of the early stage (2D growth mode) and after the QD formation (3D growth mode) are predicted in some theoretical works. [12-15] The accumulation and subsequent release of the strain in the WL not only lead to the QD formation, but also the atom migrating and segregation, which makes the WL asymmetry. Surface segregation of InGaAs/GaAs system during molecular beam epitaxy (MBE) is often described by a phenomenological model proposed by Muraki et al,[15] in which the indium content is expected to decrease exponentially towards the capping layer. The simple model agrees well with some well-established experiments.[16] But sometimes the actual WL configurations are more complex than the theoretical descriptions, leading to the formation of various kinds of self-assembled nanostructures.[17] Further more, the WL evolution process after the formation of QDs and the growth condition induced WL diversities are also interesting topics. Systematical investigations are necessary for a detailed understanding of the WLs in the SK mode.

Droplet epitaxy (DE) is another growth technique for the self-assembly of epitaxial nanostructures. Take the InAs nanostructure for example to illustrate the growth process, indium is first deposited on the buffer layer to form droplets and then react in the arsenic atmosphere for crystallization. Due to the different diffusion and crystallization rates of the deposited elements, various kinds of nanostructures can be evolved from the droplets. [18] The formations of the nanostructure in DE mode are more complicated than in SK mode and till now a universal model is still absent. Various shapes of nanostructures, such as dots, disks and rings have been fabricated with the DE mode. [19-21] High quality lattice-matched GaAs/AlGaAs quantum rings and lattice-mismatched InGaAs/GaAs nanorings, concentric quantum double rings are also obtained. [22-24] The formation of nanostructures in DE mode can be described as either strain induced (for the lattice-mismatched cases) or crystallization induced. Take the InAs/GaAs nanorings for example, the nanorings evolve from the indium droplets during the crystallization process. On the one hand, the strain induced by the lattice mismatch between InAs and GaAs can be regarded as one of the causes for the nanorings' formation, which is similar to the SK mode. [18] On the other hand, the formation of InAs rings can be also explained by the interplay between the migration of surface In adatoms and the different rate of crystallization occurred inside and at the edge of the droplets. [24]

The actual situation may be more complex than the theoretical descriptions. Unlike the SK growth mode, in which the WL thickness is determined by the lattice mismatch between the substrate and epitaxial material, the WL thickness can be varied within a relatively large range in the DE mode. There are also reports on the InAs QDs without WL grown with DE mode. [25] It is still debatable about the general existence of the thin transitionary film (the so called wetting layer) below those nanostructures. [24, 26] The configurations and evolution processes of WLs obtained with different growth conditions are also significant and interesting topics. A systematical investigation of the existence and the actual evolution processes of WL with the DE mode are necessary for an in-depth understanding of the DE nanostructure systems.

Since the WLs are embedded under the nanostructures and illuminate weakly, it is difficult to obtain the WLs related information through conventional QD characterization methods, such as atomic force microscopy (AFM) and photoluminescence (PL). Most of the WL related information is obtained through the reflectance difference spectroscopy (RDS), high resolution transmission electron microscopy (HRTEM) and cross-sectional scanning tunneling microscopy (XSTM). Both the WL configurations and the evolution processes with different growth methods are investigated systematically during the recent years. In this paper, various kinds of methods used for the investigation of the WL configurations are briefly reviewed. Some recent works in which RDS is applied to investigate the evolution processes of WLs in different growth stages of the SK mode are summarized. RDS can be also applied to determine the critical thickness of the WL. Then the effects of InAs deposition temperature and growth interruption are discussed. Applying RDS to study WLs in the DE mode grown nanostructures is involved at the last part.

## 2 The wetting layer characterization methods

Generally, methods used for wetting layer characterization can be considered as two types. One is direct imaging methods including high resolution transmission electron microscopy (HRTEM) and cross-sectional scanning tunneling microscopy (XSTM),[16, 27-30] the other is spectroscopy methods such as RDS, PLE, and modulated reflectance spectroscopy.[31-35]

### 2.1 Direct imaging method

The arrangement of individual atoms in the WL can be observed directly by HRTEM and XSTM with a high resolving power. The WL configurations such as width, substance content and segregation or the strain distribution can be obtained by analyzing the micrographs or fitting a certain measured profiles. Lattice mismatch induced strain distorts the measured lattice plane from an ideal one, which enable us to estimate the strength of strain effect, and further the concentration profiles of a certain component by finite-element calculations.[16] The atoms can be even counted directly from the high-quality micrographs to know the segregation profiles.[29] Those intuitive results are often used to estimate the accuracy of the analysis results from other WL characterization methods. But the complicated experimental setup and inconveniences of the ultra high vacuum requirement limit their further applications.

**2.2 Spectroscopy method**

Applying spectroscopy methods to examine the WLs is not intuitive but convenient. Transition energies of electrons and holes are determined by the WL configurations (composition, thickness, segregation, et al). Those transition signals can be obtained through various optical spectroscopy such as reflectance difference spectroscopy (RDS), photoluminescence (PL), photoluminescence excitation (PLE), or photo reflectance (PR). Some calculations can bridge the gap between the optical signals and wetting layer configurations.

Emission from the WL, which is an ultra thin QW (less than 3 nm), can not reach a high excitation in a simple PL measurement. [32] Take the InAs/Ga(In)As QD system for example, a redshift of the transition energies of the heavy hole exciton in WLs is found with an increasing indium deposition amount before the QD formation. [36] Since the excited carriers are mainly trapped by QDs and recombine through the QD states, the WL related energy states are difficult to be detected directly in PL spectra after the QD formation. But they may serve as carrier channels to enable the redistribution of thermally excited carriers among QDs, and further modified the temperature dependence of the optical properties of the QD system. A fast temperature-induced redshift of the emission energies and a temperature driven reduction of the PL full width at half maximum (FWHM) are commonly observed in the SK grown QDs, which are also known as anomalous PL temperature behavior. [37] By analyzing the temperature dependent PL spectra, the simulated transition energies of the carrier channels are close to those of the WLs. [38] Even the Surface-mediated indium atom migration on the patterned substrate is revealed with this method. [39]

A resonant excitation (PLE) and a low temperature are necessary to probe both the HH and LH related transition. [33] Hugues et al. determined the wetting layer thickness and composition by PLE measurement [33]. They resolved the signals related with heavy ($e_1$-$hh_1$) and light ($e_1$-$lh_1$) hole transitions of the WL. With a square quantum well (QW) assumption, different sets of composition and thickness corresponding to the same $e_1$-$hh_1$ transition energy were determined with one band k·p method. Then the light hole state position with those calculated WL configurations (thickness and In content) was determined by a more complicated k·p calculation, which took into account the coupling between valence bands and the spin-orbit splitting. Since the light hole state position changed monotonously with the WL configurations with the same $e_1$-$hh_1$ transition energy, the WL thickness and composition could be determined unambiguously. One problem is that the calculation could not take into account In segregation in the WL, which would break the square QW assumption. But a TEM measurement confirmed that the homogeneousness approximation for this calculation was accurate enough. [33]

Based on the $e_1$-$hh_1$ and $e_1$-$lh_1$ transition energies, the relationship between the WL configurations and the WL related transition energies can be also calibrated by introducing reference samples.[34] It is a more convenient method than the tedious calculations for a series of samples with the same growth conditions except the In deposition amount. The WL can be considered as an asymmetric quantum well with unknown InAs amount and segregation profile. The indium atoms in WL are usually considered to decrease exponentially from the interface of the WL and buffer layer:

$$X_{In} = \begin{cases} 0 & z < 0 \\ \dfrac{t_{WL}}{l}\exp(-z/l) & z > 0 \end{cases}$$

where z=0 is the position where InAs is deposited, $t_{WL}$ is the amount of InAs in the WL and $l$ is the segregation length of the indium atoms. Sometimes the segregation coefficient R, given by R=exp(−$a_c$/2l) with $a_c$ of the lattice constant of GaAs, is also used to indicate the segregation effect. The $e_1$-$hh_1$ and $e_1$-$lh_1$ transition energies are determined by the heavy hole (HH) and light hole (LH) energy levels in the WL and the exciton binding energies. By deciding the HH and LH exciton binding energies of some reference samples with different In content in the WLs, the exciton binding energies as functions of $t_{WL}$ can be determined with interpolation. The HH and LH energy levels are determined by the In distribution profiles of the WL. So $t_{WL}$ and $l$ can be fitted from the experimentally obtained $e_1$-$hh_1$ and $e_1$-$lh_1$ transition energies.

WL can be considered as an asymmetric quantum well (QW) due to the indium segregation effect. The symmetry reduction induces the WL related in plan optical anisotropy (IPOA), which can be characterized by RDS.[40-42] RDS is powerful and convenient to capture the WL related signals of HH and LH related transitions, and further determine the substance content and segregation coefficients. The evolution behaviors of WLs with different growth parameters, including the amount of indium deposition, temperature and growth interruption, can also be obtained by extracting the WL related IPOA signal from a series of samples.

**2.3 Reflectance difference spectroscopy**

Reflectance difference spectroscopy (RDS) is a spectroscopic technique which measures the difference in reflectance (Δr) of two orthogonal directions in the sample surface plane normalized to the incident direction under the conditions of near-vertical incidence. It is often normalized to the mean reflectance (r):

$$\frac{\Delta r}{r} = \frac{2(r_x - r_y)}{(r_x + r_y)}$$

Where r stands for the complex Fresnel reflection amplitude. The typical setup of RDS is shown in figure 1. The polarization of incident light and the modulation axis of photoelastic modulator (PEM) are arranged to form an angle of 45 degree with the crystal axis with optical anisotropy ([110] and [1-10] for samples with a square symmetry). The reflected light pass through the photoelastic modulator (PEM), analyzer, and finally reach the detector. If reflection coefficients along the two optical axis are the same, the detected signals do not contain the modulated signals from PEM. Otherwise, the relative difference of reflectance along the two optical axis (Δr/r) can be obtained directly from the PEM modulated part.

As a non-destructive optical probe technique, RDS is capable to operate at a wide range of environments and material systems.[43-45] It is widely used in the MBE and MOCVD growth processes as a kind of in-situ monitor to control and optimize the technological parameters.[46-48] Time resolved RDS measurement at 2.6 eV and 4.2eV can resolve the self-assembled InAs/GaAs QD formation induced surface changes, including the WL growth, 2D-3D transition and surface re-smoothing during the capping process.[49, 50] As a static spectroscopic method, RDS is also widely applied to investigate the surface and interface of semiconductor materials, such as the chemical bonds and reconstructions of the semiconductor surface, the surface and interface electronic states in the region of the Fermi level .[51-54] Features in RD spectra can be identified with electronic transitions between surface and interface states, which makes RDS a sensitive tool to probe the surface state behavior.

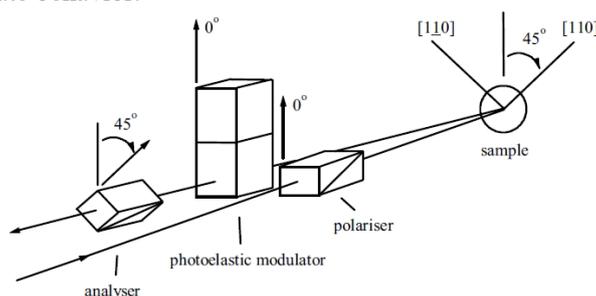

Figure 1: A schematic arrangement of the RDS

Recently, the potential of RDS in investigating the in plan optical anisotropy (IPOA) of III-Vsemiconductor quantum wells and supper lattices is illustrated by the works of Chen and co-workers.[55-60] An inverse linear dependence on the well width of the IPOA in the single quantum well (QW) GaAs/Al$_x$Ga$_{1-x}$As structure is revealed, and the interface-related contribution to optical anisotropy in the symmetric and asymmetric single QW and supper lattices is investigated systematically. Generally speaking, the asymmetry of QW can be considered as bulk like or interface like. Bulk like asymmetry can be caused by an electric field or compositional variation across the QW, while interface like asymmetry may raised from the difference in interface bonds, interface composition profile (segregation effect), the anisotropic strain, etc. If no external electric field is applied and the build-in electric field is very weak, the electric field induced asymmetry can be neglected. Besides, the differences of the interface bonds in the common-atom QW, like Al$_x$Ga$_{1-x}$As/GaAs and In$_x$Ga$_{1-x}$As/GaAs, are expected to have a weak IPOA contribution for the lack of intrinsic nonequivalence of two interfaces. Therefore, the IPOA of a common-atom QW is mainly originated from the segregation and anisotropic strain effect in the QWs.[61] The segregation and other WL configurations can be determined by fitting the RD spectra. WLs in self-assembled structures can be considered as asymmetry QWs due to the build-in strain and segregation effect. RDS is a powerful tool in detecting the WLs related IPOA and determining the WL configurations. It has been also applied to systematically investigate evolution of the WLs obtained through different growth modes.

### 3 Wetting layer evolution in SK mode grown QDs

#### 3.1 A general description of the wetting layer evolution in SK mode

By examining the WL configurations in a series of InAs/GaAs quantum dot systems with a certain indium content gradient, one can analyze the wetting layer evolution process. For the MBE growth technique, such kind of samples could be obtained by stopping the substrate rotating during the InAs deposition.[62] As shown in the inset of figure 2, if the incidence angle of the indium beam and the distance between the indium source and the sample centre are known, an almost linear variation of the amount of InAs with the wafer position can be predicted on the basis of the cosine law for the MBE source beam.[63] The curve shown in figure 2 corresponds to a nominal In deposition amount of 2ML. In order to study WLs with different deposition thicknesses under the same growth condition, the wafer can be cut into several pieces after growth and each piece stands for a typical indium deposition thickness.

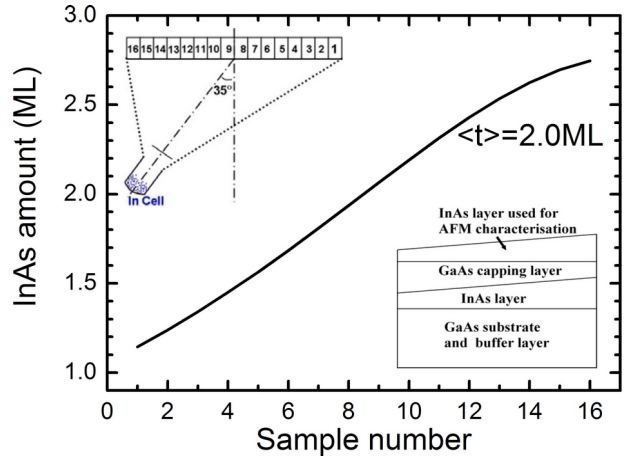

Figure 2: Deposited InAs amount among the 16 pieces uniformly cut from 2-inch wafers predicted by the cosine law. The insert shows schematically the slant incidence of indium beam and the InAs layer with a gradually changed thickness.[64]

### 3.1.1 WL evolution in 2D and 3D growth stages

Chen and his coworkers investigated the evolution of indium content in the WL of an InAs/GaAs quantum dot system systematically with RDS.[34, 35, 65] The InAs layer is deposited at 510°C, and with a deposition rate of 0.1ML/s to form QDs. By stopping the substrates rotating, a gradually changed InAs thickness (with a nominal deposition amount of 1.9ML) is achieved and cut into 16 pieces to be examined with AFM and RDS.

The inset of figure 3 shows the AFM images (0.5μm×0.5μm) of samples 6 and 12, where InAs quantum dots are clearly observed on the surface. Figure 3 shows the shift of QD density with an increasing deposition amount by analyzing the AFM images. The QDs start to appear from sample 6 (corresponding to a nominal InAs deposition amount of 1.6ML), indicating that the 2D growth mode (samples 1–5, with a nominal InAs deposition amount ranging from 1.09ML to 1.49ML) is replaced by 3D growth mode. For samples 6–13 (with the InAs deposition amount from 1.60ML to 2.41ML), the dot density increases quickly and saturates at a level of about $3.5\times10^{10}cm^{-2}$, while the average QD diameter and height change slightly (not shown). For samples 14–16 (InAs deposition amount from 2.50ML to 2.61ML), the dot density increases again with a decreasing QD diameter and height, which may due to the thermal gradient at the edge of the wafer.

The RD spectrum of sample 6 is shown in figure4. Three features at 876.5, 895.7 and 924.1 nm are clearly observed and assigned to the transition signals of GaAs band edge, LH and HH transitions in the WL, respectively.[34] Figure 5 gives the RD spectra of all the 16 samples. Triangles, squares and circles denotes GaAs band edge, LH and HH transition related signals. The LH and HH transition energies, which are mainly affect by the indium content of WLs, decrease linearly at first (samples 1-5) and almost keep constant hereafter (samples 6-16). Briefly speaking, it indicates that In content in the WL increases linearly during the 2D growth process and keeps an upper-limit value during 3D growth. The change of energy spacing between the LH and HH transition energies may come from the differences of indium segregation degree among all the samples.

By introducing reference samples and a linear interpolation, the exciton binding energies are determined. The relationship between the WL configurations and the WL related transition energies is calibrated with the method mentioned in previous part. The calculated WL thickness and the segregation coefficient can be seen in figure 6. Distinct differences of the evolution processes of the In content and segregation effect in WL between the 2D and 3D growth stage can be easily figured out. In the 2D growth stage, both the In content and the segregation coefficient increase linearly with an increasing In deposition amount. It indicates that all the newly deposited InAs is incorporated into the WL, and the segregation effect is enhanced by the increase of In amount in the WL. As to the 3D growth stage, indium content in the WL is almost saturated. A distinct decrease of the segregation coefficient can be attributed to the QD formation induced strain relaxing in WLs.

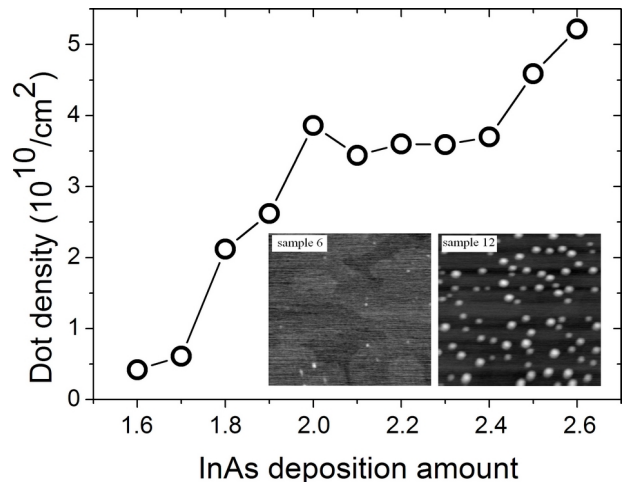

Figure 3: Averaged densities of QDs from AFM. The inset pictures are the AFM images (0.5μm×0.5μm) of samples 6 and 12[34]

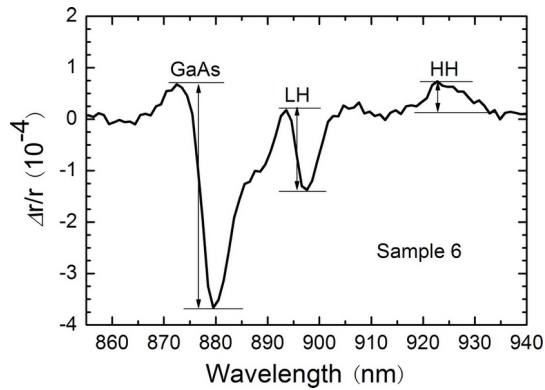

Figure 4: RD spectra of sample 6 (the nominal InAs deposition amount of 1.60ML). The wavelengths and RD intensities for the GaAs band edge, LH and HH related transitions of the WL are indicated by the vertical lines with arrows [34]

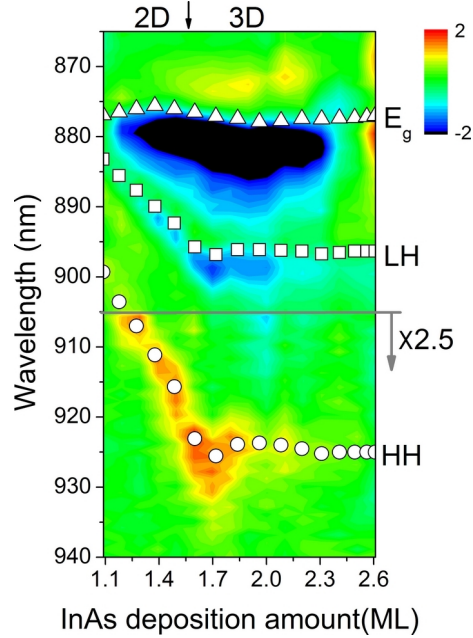

Figure 5: The RD spectra of samples with InAs layer deposited at 510°C and a nominal InAs deposition amount of 1.9ML. For clarity the spectra in the long wavelength range (>905 nm) are magnified 2.5 times[34]

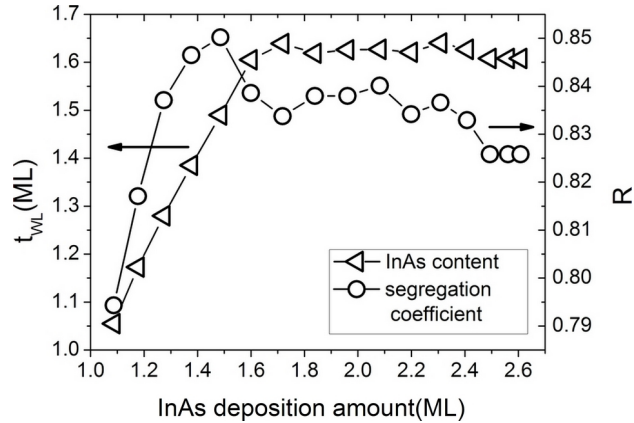

Figure 6: The calculation results of the InAs content in the WL $t_{WL}$ (triangles), and the segregation coefficient $R$ (circles) for samples with a nominal InAs deposition amount of 1.9ML[34]

### 3.1.2 WL evolution in the ripening process

An increasing deposition of InAs after the formation of stable QDs may lead to a ripening process.[66] When the average size of QDs reach a certain critical size, dislocation may be generated to relax the strain energy, accompanied with a fast growth of the dislocated QDs.[67] As shown in figure 7, some huge dots appears corresponding to a large In deposition amount. A particular WL evolution process may be also involved.

Based on a series of samples (16 pieces) with InAs layer deposited at 530°C, a deposition rate of 0.008ML/s and gradually changed InAs amount ranging from 1.14ML to 2.75ML, Chen and his coworkers investigated the WL evolution in the QD ripening process.[35, 65] As shown in figure 8, different growth mode can be distinguished clearly from the AFM analyzing result. Compared with the general 3D growth stage, a smaller normal QD density, whereas a larger huge dot density was found in samples12-16 (with the nominal InAs deposition amount ranging from 2.43ML to 2.75ML), to which a ripening process could be assigned. The WL related signals were observed in all the 16 samples from RD spectra shown in figure 9. The evolution of indium content and segregation in WLs were calculated based on the LH and HH transition energies, which are shown in figure 10. Although more InAs is deposited, the indium content in the WLs decreases in the ripening process, corresponding to a blue shift the LH and HH transition energies in figure 9. This can be attributed to the formation of the huge dots and dislocations, which can serve as sink centers for not only the newly deposited InAs, but also the indium atom in the WL.

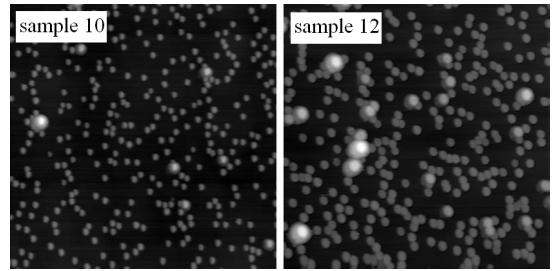

Figure 7: Some large dots appear in the ripening process revealed in the AFM images[35]

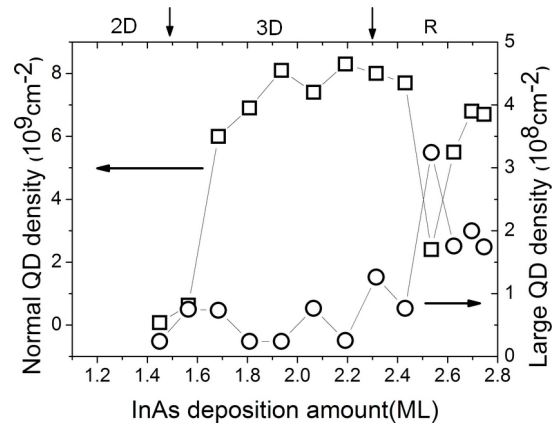

Figure 8: Density of InAs dots $N_{QD}$ (squares), density of huge InAs dots $N_{LQD}$ (circles) obtained from the AFM results. [35]

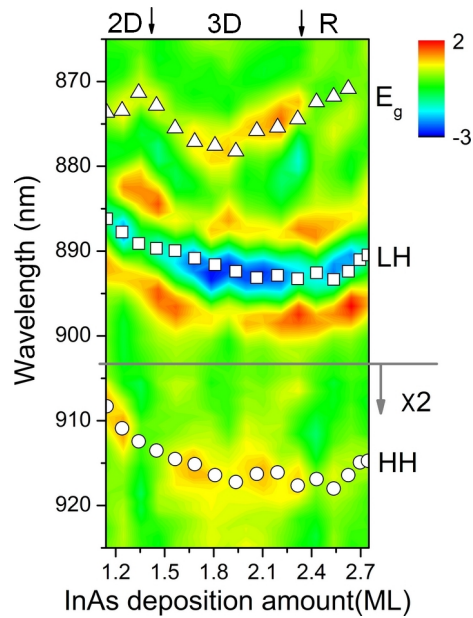

Figure 9: $d^2\rho/d\lambda^2$ spectra of samples with the InAs layer deposited at 530°C and 0.008ML/s. The nominal InAs deposition amount is 2.0ML. The wavelengths of the GaAs band edge and WL related transitions are indicated by

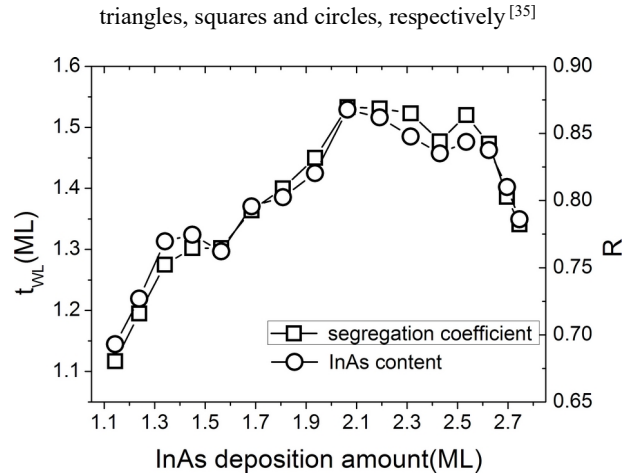

triangles, squares and circles, respectively [35]

Figure 10: The deduced InAs content in the WL $t_{WL}$ (circles) and the segregation coefficient R (squares) for the 16 samples with a nominal InAs deposition amount of 2.0ML. [35]

### 3.1.3 QD formation induced spectral changes

Generally, the QD related information is absent in the RD spectra. A comparison between RDS and other reflectance based optical characterization methods, such as reflectance spectrum and photo modulated reflectivity measurement (photoreflectance), is interesting. Photoreflectance is also sensitive to the ultra thin WL layers embedded in the matrix. The QD related signals can be also discerned with clear features. [32, 68] Some useful discussions on WLs in the self-assembly structures have been carried out. As a relatively less sensitive measurement, the reflectance spectrum is hard to catch the single quantum well or WL related signals. But in some exceptional cases that the WL related features are discerned in the reflectance spectra, some interesting results have been observed when they are compared with the results from RD spectra.

Figure 11 gives the RD spectra and reflectance spectra of samples with a gradually changed InAs deposition amount in mapping style. The InAs layer is deposited at 520℃ with a nominal thickness of 2ML. Figure 12 gives a comparison of the WL related transition energies determined with RD spectra and reflectance spectra. After the QD formation (corresponding to the critical thickness of 1.6ML), the transition energies determined with reflectance spectra show clear blueshifts (50meV at the maximum) compared with the results obtained with RD spectra. The deviation may originate from two kinds of WLs whose differences in thickness are the order of sub-monolayer. The thicker WLs with lower transition energies have a strong IPOA but a small density of state, leading to the strong RD signals but negligible reflectance signals. Whereas the thinner WLs with higher transition energies are isotropic and can not be reflected from RD spectra, but they may contribute a lot to the reflectance spectra. A detailed description and its relation with the QD formation is still under investigation.

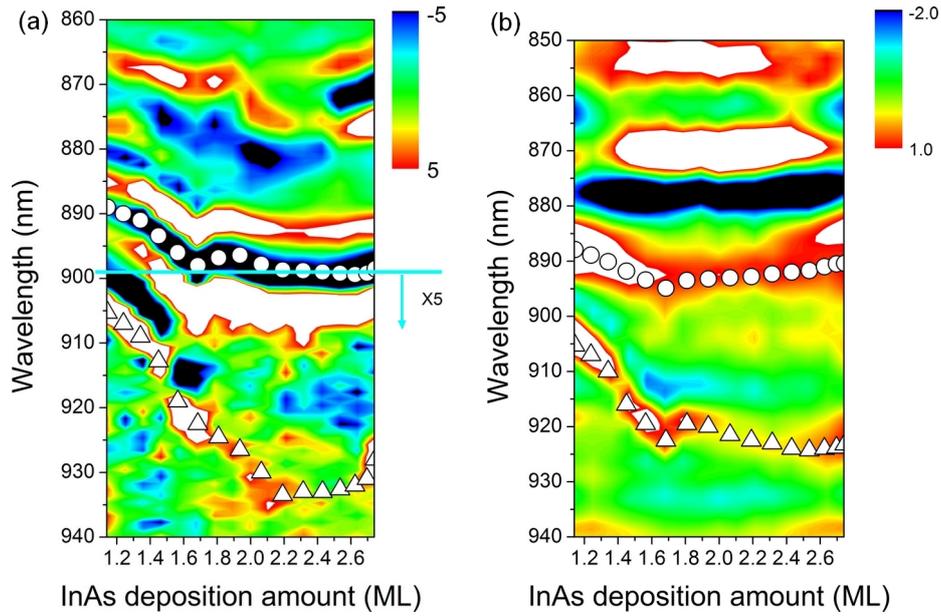

Figure 11: (a):$d^2\rho/d\lambda^2$ spectra of samples with the InAs layer deposited at 520℃. The nominal InAs deposition amount is 2.0ML. (b):dlnR spectra of the same series of samples. The wavelengths of the LH and HH transitions are indicated by circles and triangles, respectively

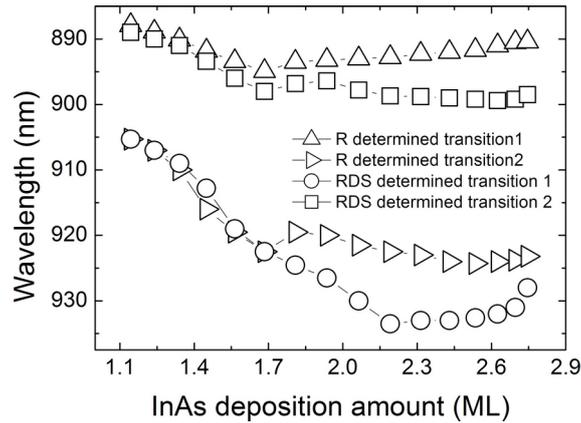

Figure 12: Different WL related transition energies determined with RDS and reflectance spectrum for samples with gradually changed InAs deposition amount

### 3.2 Segregation effect

Generally speaking, segregation in WL is more significant with increasing InAs deposition amount, but varies a lot among different samples, as shown in figure 6 and figure 10. An insight of the segregation effect can be obtained from the relationship between the segregation coefficient and the WL thickness.[35] As shown in figure 13, segregation coefficient varies linearly with the WL thickness for the sample with a higher InAs layer deposition temperature (530℃). Whereas for the sample with lower InAs deposition temperature (510℃), a linear relation at the initial stage, then a significant deviation is found. The linearly determination of the segregation coefficient by the WL thickness (InAs amount) means that strain is a driven force for the In segregation in the WL. The more strain is accumulated; the more severe segregation is found. The formation of QDs, which relaxes strain in the WL, can greatly affect the segregation of indium atoms in the WL. Further, the average segregation coefficient of the sample with InAs layer deposited at 530℃ is smaller than the segregation coefficient of other sample in figure 13. It can be explained that the higher growth temperature enhances atomic mixing and relaxes strain in the WL, so the segregation effect is also weakened.

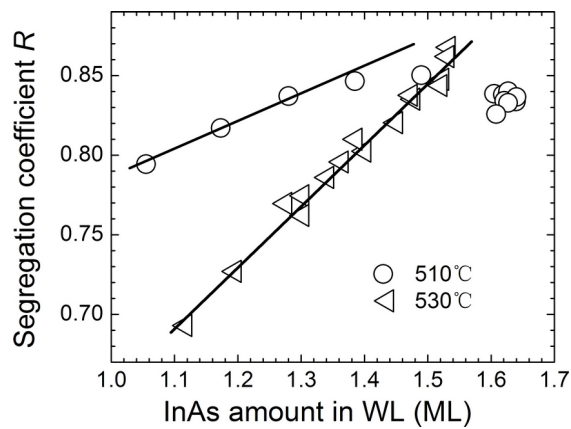

Figure 13: Segregation coefficient versus InAs amount in WL for the samples with InAs layers grown at 510°C and 530°C, respectively

### 3.3 RDS in determining the critical thickness

RDS also provides a new way to decide the critical thickness of the 2D-3D growth transition.[64] As to the In(Ga)As/GaAs system, in-situ RDS monitor of the 2D-3D transition has been reported at energies around the $E_1$ and the $E_1+\Delta_1$ transitions.[69] The critical thickness can be also discerned from an abrupt change of the WL related transition energies obtained from RD spectra.

As shown in figure 14 (a), different growth stages can be discerned clearly from the shifts LH transition energy with the change of In deposition amount. In the 2D growth stage, the newly deposited InAs are all incorporated into the WL corresponding to a linear redshift of the WL related transition energy. In the 2D-3D transition stage, the thickness of the WL reach the critical thickness and the QD formation triggers an abrupt change of the migration and distribution of the indium between the WL and QDs. Above the critical thickness, the newly deposited InAs are largely incorporated into QDs and the WL related transition energies continue to redshift but with a smaller slope or even keep constant. In figure 14 (a), The QD formation triggered kinks around 1.68ML (sample 6) can be seen in all the samples with different InAs grown temperatures, which can be used to determine the 2D-3D transition or the critical thickness. At the same time, InAs in the WL is saturated at the critical thickness, corresponding to a rapid change of the anisotropic strain. An abrupt change of the IPOA is also observed at the critical thicknesses corresponding to the kinks of LH transition energy (figure 14 (b)).

In some situations, the WL critical thickness determined by RDS can be more authentic than the commonly used methods, such as RHEED in-situ observation and AFM simulating, which may be affected by the thermodynamic fluctuation induced pre-QDs or quasi-QDs.[64] The inset of figure 14 (a) shows the critical thicknesses of samples with different InAs layer deposition temperature determined by RDS and AFM. A significant deviation can be found for samples with the InAs layers grown above 510 °C. The critical thickness determined by AFM is generally considered as a kinetic aspect,[70] whereas the RDS results based on the WL evolution is an free-energy based analysis. Some small dots with a low density may appear below the thermodynamic critical thickness due to the fluctuation. While a large number of dots are evolved from the WL when the thermodynamic conditions are meet. The desorption of indium atoms may also influence the QD formation. Those factors make the AFM results deviating from the RDS results. RDS can conveniently reveal the critical thickness determined by the thermodynamic condition.

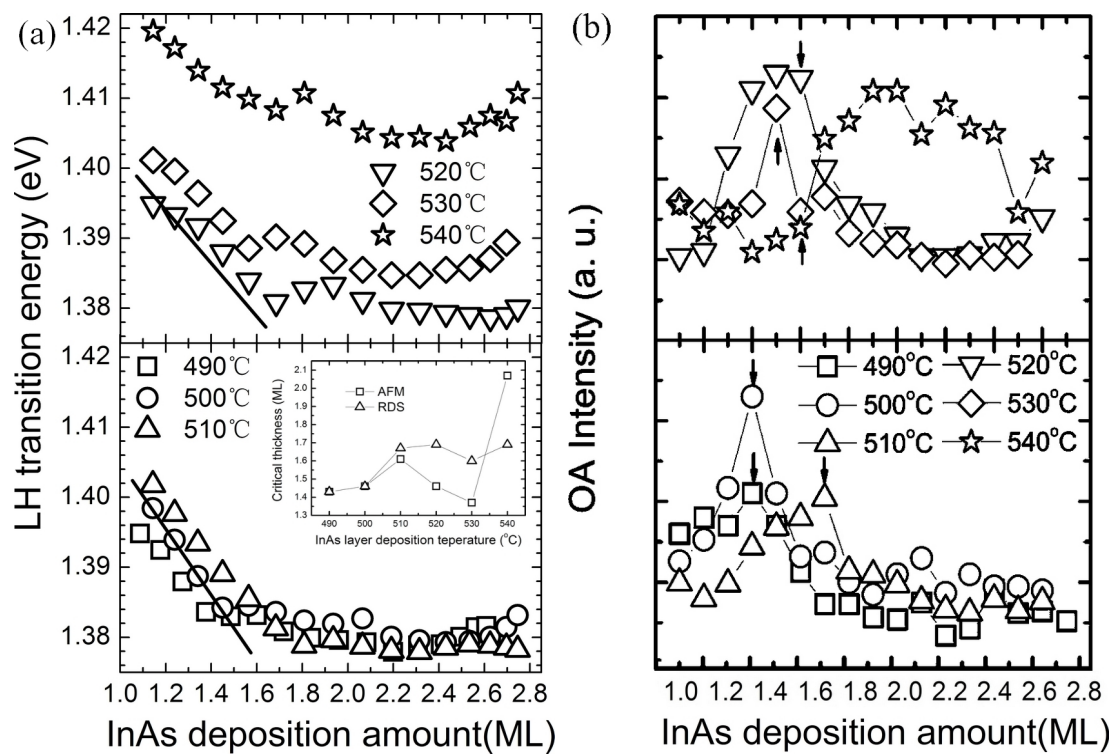

Figure 14: (a) LH transition energies (b) intensities of IPOA of WL for six samples grown at different temperatures. The inset figure of (a) shows the critical thicknesses of samples with different InAs layer deposition temperatures determined with RDS and AFM. [64]

The intensities of in plan optical anisotropy (IPOA) obtained from RD spectra can be also used to determined the WL critical thickness. Generally speaking, IPOA originates from symmetry reduction of the WL, which may arise from the segregation and the lateral compositional modulation (LCM). The segregation effect in the WL has been discussed previously, which can be attributed to the strain driven process. The LCM in WLs can be considered as the results of surface reconstruction and anisotropic adatom diffusion.

For InAs/GaAs system, it is well known that the two-dimensional stripes and pits are existent in WLs and elongate along [1-10] direction, which lead to anisotropic strain in the WL and IPOA detected by RDS. Evolution of those structures may be closely connected to the formation of QD. In other words, anisotropic structures and strain accumulate during InAs deposition process until the formation of QDs, then quickly transform into QDs, which lead to the largest IPOA intensities at the very beginning of the QD formation. This phenomenon has been verified by many experiments, indicating that the evolution of the IPOA intensities can be also used to determine the WL critical thickness. As shown in figure 14 (b), evolution of the IPOA intensities agrees well with the evolution of WL related transition energies for sample with different InAs deposition temperatures. The emergence of the IPOA maximum can be also used to determine the WL critical thickness. [64] Similar evolution process of the IPOA intensities which are originated from the anisotropic strain in GaAs below InAs layer is also found. As shown in figure 15 (b), both the strain and the IPOA intensities reach to the maximum at the critical thickness.[71] But some exceptions are also found, such as the IPOA intensity shows an upsurge at the critical thickness and keeps fixed during the SK stage [65], then suffers a decrease if some large dots appear. The intrinsic characters of those anisotropies in WLs and their relationships with the formation of QDs are still under investigation.

Abrupt changes around the critical thickness can be also found in the GaAs buffer layer related signals. As shown in figure 15, for samples with exactly the same growth conditions but different InAs deposition amounts, abrupt changes of the GaAs band edge transition energies and their IPOA intensities are found at the deposition amount of 1.6ML, which is believed as the critical thickness. [71] Tensile strain exists in GaAs due to a smaller lattice constant, which lead to a redshift of the transition energy until the critical thickness. Strain in GaAs can be also estimated from the changes of transition energies. The hydrostatic strain can be calculated with $\varepsilon_H = \Delta E_0/(a_c - a_v)$, where $\Delta E_0$ is the energy shift and $a_c$ ($a_v$) is the deformation potential of the conduction (valence) band. The maximum value of the tensile strain can be determined as $2.4 \times 10^{-3}$ corresponding to a InAs deposition amount of 1.6ML, which is much smaller than the misfit strain of 0.07 between GaAs and InAs. So the misfit strain is mainly accommodated by the deposited InAs.

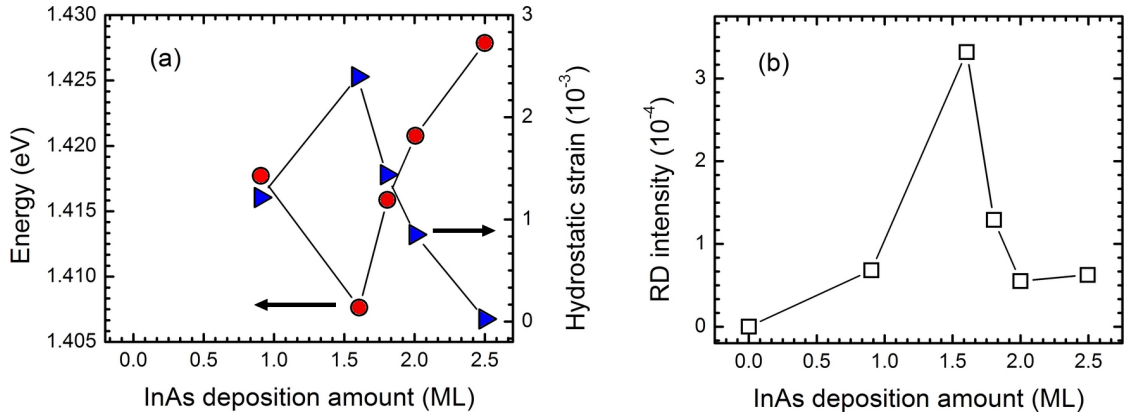

Figure 15: (a) GaAs band edge related transition energies (solid squares) and the averaged hydrostatic strain of GaAs surface (open squares), (b) RD intensity as functions of the InAs deposition amount. [71]

### 3.4 Desorption of indium atoms

The grown temperature of InAs layer can affect the WL evolution and critical thickness greatly. [64] As shown in the inset of figure 14 (a), the WL critical thickness varies with InAs deposition temperature intricately. Various kinds of results have been shown in previous studies.[72, 73] An explicit conclusion is still necessary.

Generally speaking, a kinetic description of the InAs transfer for a MBE grown sample can be described with figure 16. It take a certain period of time for the newly deposited InAs to form the equilibrium WLs, corresponding to the WL formation rate $r_{eq}$. The InAs desorption process also exists due to the relatively high deposition temperature. Indium desorption in different temperatures during MBE growth of the strained InAs layer is generally considered in the Arrhenius form with a desorption rate $r_{des} \propto \exp(E_{des}/k_BT)$, where $E_{des}$ is the activation energy for indium desorption processes and $k_B$ the Boltzmann constant. Above the critical thickness, the formation of the QDs consumes the newly deposited InAs via the WLs. The consumption of indium atoms from WL by the QDs can be also described in the form of thermal activation with a rate $r_{QD} \propto \exp(E_{QD}/k_BT)$, where $E_{QD}$ is the activation energy for the stress-driven indium diffusion from WL to QDs. If the WL growth rate $r_{eq}$ is larger than the sum of the indium desorption and consumption rate ($r_{des}$ and $r_{QD}$), the WL becomes thicker. Otherwise the WL becomes thinner even though InAs is deposited continuously.

WLs obtained with different deposition temperatures show different evolution processes with the indium content. As shown in figure 14 (a), For samples obtained at a relative low temperature (490-510°C), the LH transition energies are almost fixed after the 2D-3D transition kink, while for samples obtained at high temperature (520-540°C), the LH energies show a clear overshoot before stabilization. Since the WL related transition energies are mainly determined by the indium content, it is straightforward to know the evolution of the indium content of WLs with different grown temperature. The transition energies overshoot (or the indium depletion) after QD formation can be understand as the interplay of the different incorporation and consumption rates of indium in the WL. The WL growth rate is considered to be constant for a certain temperature. As shown in figure 3 and figure 8, QDs with a certain density and size may appear suddenly right above the critical thickness, corresponding to a sharp increase of the indium consumption rate $r_{QD}$. If $r_{QD}$ exceeds $r_{eq}$, the net indium in the WL decreases and the WL related transition energies show slightly blue shift. For samples with InAs layer deposited at 520-540°C, the blue shift can be seen right after the 2D-3D transition kink. The depletion of indium atoms relaxes strain in WLs, which weakens the driving force of the indium segregation from WL to QDs and reduce $r_{QD}$. If $r_{QD}$ becomes smaller than $r_{eq}$, InAs in the WL accumulates again until the next equilibrium, which lead to the overshoot of the transition energies in figure 14 (a). Absence of the overshoot for lower InAs deposition temperature can be explained as a smaller $r_{QD}$ for those samples.

For a high InAs deposition temperature, the indium desorption process is important, which reduces the WL thickness and elevates the LH transition energies. As shown in figure 14 (a), the average LH transition energy is higher for samples grown in 530-540 °C. Before the QD formation, InAs amount in the WL deviates from the deposition amount because of desorption. If the relationship of InAs content and LH transition energies is calibrated for those samples, the temperature dependent indium desorption processes can be analyzed based on figure 14 (a). The temperature dependence of indium desorption for the sample with a nominal deposition amount of 1.45ML (sample 4) in an Arrhenius plot is shown in figure 17, where the activation energy can be calculated by a linear fit. For samples 1-4 (with a nominal InAs deposition amount ranging from 1.14 ML to 1.45 ML), the activation energies are determined about 2.8eV, which are similar with the results reported previously. For other samples, a certain density of QDs appears and the amount of InAs incorporated into the QDs is necessary in order to get the indium desorption amount and activation energies. But it is clear that the desorption processes contribute to the differences of InAs in the WL among samples with different growth temperatures.

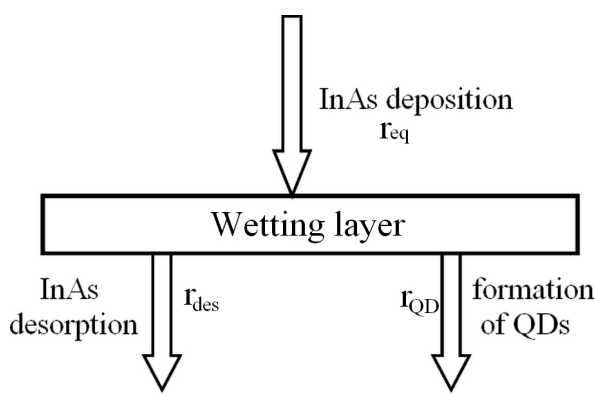

Figure 16: An illustration of the transfer of InAs via the wetting layer during InAs deposition

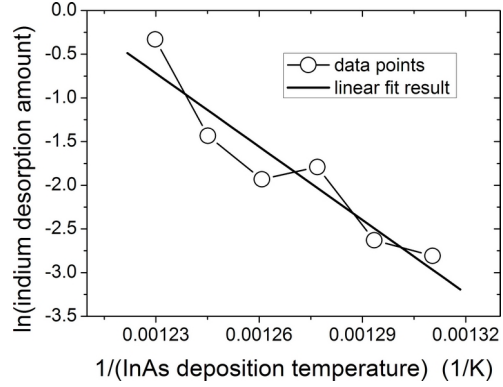

Figure17: Arrhenius plot of the temperature dependence of indium desorption amount for the sample with a nominal deposition amount of 1.45ML (sample 4)

### 3.5 Effect of growth interruption

Growth interruption (or post-growth annealing, GI) for a certain period of time can be introduced after the deposition of InAs layer to improve the structural and optical properties of QDs.[74, 75] As an important QD growth and modification technique, GI greatly impacts the WL evolution progresses, and further the QD formation and physical properties. Applying RDS, the WL evolutions with different periods of GI are investigated, which can deepen the understanding of the QD growth processes. Zhou et al. fabricated three samples with different GI (0s, 30s, 90s) after the deposition of the InAs layer, and the substrate does not rotate to achieve a gradually changed InAs amount (with a nominal deposition amount of 2ML, 1.9ML and 2ML, respectively). Figure 18 (a) shows the areal densities of the three samples by analyzing the AFM images. It is clear that the QDs are formed at different In deposition amounts for samples with different periods of GI (1.6ML, 1.37ML and 1.26ML for samples with 0s, 30s, 90s GI, respectively), which means that GI changes the critical thickness of the WL. GI also changes the shape of QD significantly (not shown here).

A two-stage evolution for the sample without GI, and three-stage evolution for the samples with 30s and 90s GI are revealed with RDS.[76] The shifts of LH transition energies of three samples are obtained with RDS and shown in figure 19, from which different growth stages can be resolved clearly. In the 2D growth stage, a linearly redshift of the LH transition with almost the same slope for the three samples can be seen, which indicates that all the deposited InAs are incorporated into the WL. The 2D-3D transitions, which correspond to the kinks of the LH transition energies (marked with arrows in figure 19), appear at different In deposition amount for samples with different GI. The RDS determined critical thicknesses agree well with the AFM results. Above the critical thickness, the SK growth stage begins and WLs of the three samples show different evolution processes. For the sample without GI, the LH transition energy, or the In content in WL remains constant, which shows a two-stage evolution. Whereas for samples with 30s and 90s GI, a continuously redshift but with a smaller slope up to a thickness of 2.2ML is followed. After that, the LH transition energies are fixed or slightly increased. It is a typical three-stage evolution in which the newly deposited InAs are partly incorporates into the QD above the critical thickness and then a ripening stage is followed. Some huge dots appear in the AFM images, which also indicate the existence of a ripening process.

A further understanding on the effect of GI can be obtained with an equilibrium growth model.[66] the strain and bonding energy, the elastic energy of edges, the stress fields of the interfaces, the extra island surface energy, and the island-island interaction are considered in this model. As shown in figure 20, $n_1$ and $n_2$ refer to the In content in the WL and SK island coverage, respectively. LH transition energies are considered to be linearly related with $n_1$ for simplification. The two-stage and three-stage evolution are discerned and agree well with the RDS experimental results by varying the extra island surface energy γ and the island-island interaction b in the model. For the two-stage evolution, γ=0.32 and b=0 are decided. The newly deposited In are all incorporate into the WL below the critical thickness (2D growth stage), and then are all absorbed by the QDs when entering into the 3D growth stage. It is a typical In migrating behavior for the SK mode grown QDs which is also referred previously. As to the three-stage evolution, γ=0.305, b=15 and γ=0.30, b=15 are decided for the samples with 30s and 90s GI, respectively. Above the critical thickness, part of the deposited In are incorporated into the WL and the rest are all absorbed by the QDs. Above a InAs deposition amount of 2.2ML, both $n_1$ and $n_2$ are fixed and the deposited In form the ripened QDs, i.e., the growth mode is transferred to the ripening stage. As to the QDs, longer GI leads to a decreasing average aspect ratio, which is confirmed by the decreasing extra island surface energy γ (reduced from 0.32 to 0.30 corresponding to an increasing GI from 0s to 90s) and the AFM analyzing results (as shown in figure 18 (b)). Further, GI enhances the interaction between the adjacent QDs. A strong island-island interaction requires a large b. For the sample without GI, b=0 indicates the absence of elastic interaction between the QDs. Whereas a certain period of GI, corresponding to b=15, enhances the elastic interaction and provides enough time for the In migrant between QDs.

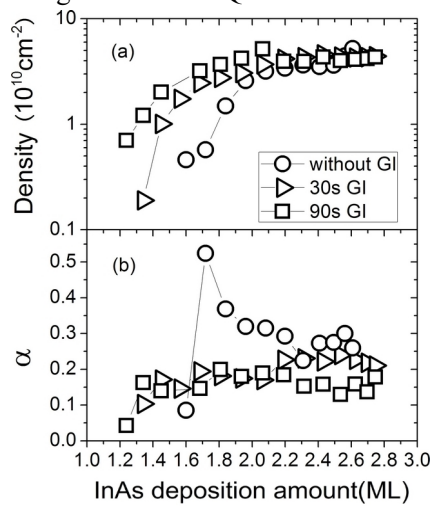

Figure 18: (a) Density, (b) aspect ratio α of InAs QDs of the three samples obtained from AFM images as a function of InAs deposition amount. The aspect ratio α is defined as the ratio of averaged height to averaged diameter.[76]

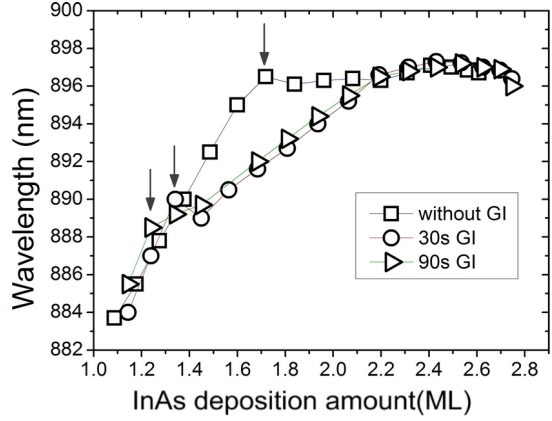

Figure 19: LH transition energies of the WLs for the three samples with different GI durations. The three arrows mark the critical thicknesses.[76]

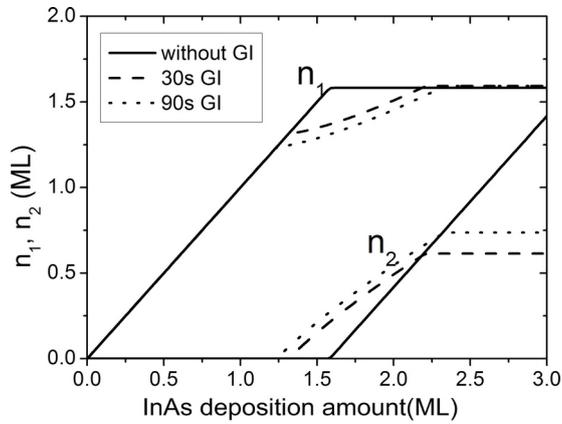

Figure 20: WL thickness ($n_1$) and SK island coverage ($n_2$) as a function of InAs deposition amount for samples with different periods of GI.[76]

## 4 Wetting layer in Droplet epitaxy mode grown nanostructures

### 4.1 Discontinuous wetting layers in droplet epitaxy mode grown nanostructures

Various kinds of the WLs below the DE mode grown nanostructures have been reported. With the use of XSTM, ultrathin WLs or even WLs with variable thicknesses are observed in the DE grown GaAs/AlGaAs system. [30, 77] There are also reports that WLs are absent in the DE grown InGaAs/GaAs nanorings. [24, 37]

The existence of the WL can be investigated with RDS. Zhao et al. detected the WL transition related signals on DE mode grown samples with higher (2.5ML) and lower (1.35ML) nominal indium deposition amount. The samples are grown with MBE. Indium is deposited with a gradually changed amount by stopping the substrate rotating at 120°C, and then $As_4$ flux with a beam equivalent pressure of $1.35 \times 10^{-5}$ Torr is supplied to the sample surface for several minutes to form the nanostructures. The WL related transition signals are all detected on these samples with RDS. Figure 21 shows a typical RD spectrum. The features around 1.35eV, 1.38eV, and 1.42 eV are assigned to the HH and LH related transitions of the InGaAs WL and GaAs band edge, respectively. The transition energies are similar to those observed in InAs/GaAs QD system grown with SK mode. The RDS result demonstrates that there is an In(Ga)As WL below the nanorings on those samples. For samples grown at 480°C, the WL related signals are absent in the RD spectra, which indicates that there is no WL on samples obtained at high temperature.

It should be argued that although the photoluminescence of the nanostructures does not show the typical temperature properties of the SK grown QDs, the WLs may still exist. [37] For samples grown at 120°C with a nominal indium deposition amount of 2.5ML, the PL spectra do not show the temperature-induced fast redshift and other anomalous behaviors. A temperature dependent photoluminescence combined with a rate-equation-based analysis reveals that the transition energy of the carrier channels is 1.575 eV, which is impossible for a continuous InGaAs WL but rather the compressed GaAs bulk material. Therefore, the nanorings are directly connected with the GaAs substrate. The In(Ga)As WLs detected in the RD spectra are discontinuous and may depleted around the InAs nanostructures,[78] which is impossible for the SK grown QDs.

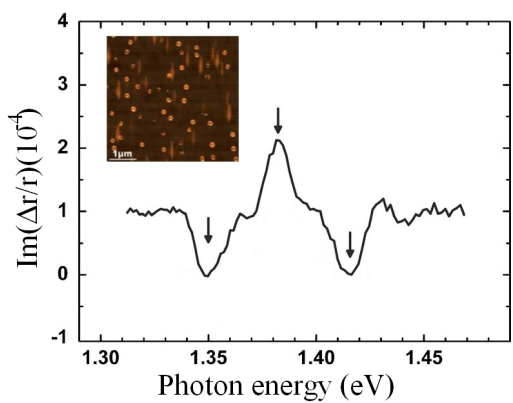

Figure 21: The RD spectrum measured at room temperature for the DE grown nanorings. The HH, LH, and GaAs features are indicated by arrows from left to right. The inset shows the AFM image of InAs structures for the as-grown samples.[78]

### 4.2 Wetting layer evolution in droplet epitaxy mode grown nanostructures

There is still lack of theoretical models to generally describe the WL evolution in the DE mode. But some basic principles have been revealed by some recent experiments.

The WL evolution processes of the two series of samples grown at 120°C with different nominal indium deposition amount are investigated with their RD spectra. [78] Figure 22 and figure 23 show the AFM and RDS result of those samples. For samples with a nominal indium deposition amount of 1.35ML, different critical thicknesses for the droplets and nanostructures are found. [79] As shown in figure 22 (a), when the indium deposition amount exceeds 1.1ML, QDs with an areal density of $10^8$ cm$^{-2}$ form to relax the mismatch strain. The droplet and further the nanorings (the inset of figure 21) appear when 1.4ML indium is deposited. As to the samples with a nominal indium deposition amount of 2.5ML, nanorings appear at all samples because the deposition amounts for these samples all exceed 1.4ML.

The WL related HH and LH transition signals can be discerned in all the samples. As shown in figure 22, for the samples with relatively lower indium amount (a nominal thickness of 1.35ML). The intensities of IPOA become stronger with an increasing indium deposition amount. The LH related energies show continues redshift at first but slightly blueshift when the density of the nanorings is saturated from the AFM analysis. The amount of indium in the WL can be linearly related to the HH and LH transition energies. Combined with the AFM analyzing results, which reveal the evolution of the nanoring densities, the indium migration behavior with an increasing deposition amount can be analyzed through the shift of WL related transition energies in RD spectra. As to the samples with a nominal indium deposition amount of 2.5ML (figure 23), the deposited In is incorporated both the droplets (nanorings) and the WLs for samples with an indium amount ranging from 1.43ML to 2.10ML (samples 1-6). For further deposition (samples 7-11, with an indium amount from 2.26ML to 2.89ML), a part of indium atoms migrate into the droplets (nanorings) and the rest into the WL, corresponding to a slower redshift of the transition energies in RD spectra. For samples 12-16 (with an indium amount from 3.04ML to 3.43ML), a decrease of the In amount in the WL can be revealed from the blueshifts of both the HH and LH transition energies, while the densities of the nanorings stay almost unchanged. For these samples, both the newly deposited indium and indium atoms in the WLs may be absorbed by dislocations induced by the formation of some large dots (not shown here), which is similar with the ripening stage in SK growth mode.

Some distinct differences of the WL evolution processes of the InAs/GaAs system between the DE mode and the SK mode can be revealed from RD spectra in figure 23(b). For the SK mode, the energy of the WL transitions can be limited to a very narrow dispersion range after the QD formation, corresponding to a stable WL. While in the case of DE mode, the energy of the WL transitions disperse in a wide range after the ring formation and stable transition energy is hard to be obtained. This is because the WL thickness is primarily determined by the lattice mismatch in SK mode, whereas it is also affected by the behaviors of the droplets in DE mode.

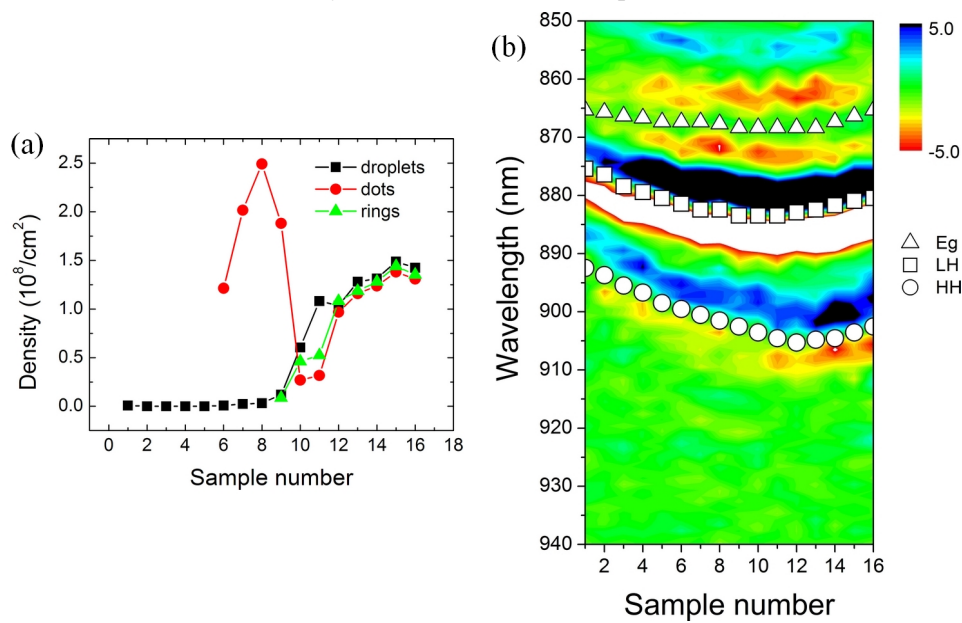

Figure 22: (a) The areal density of InAs QDs, droplets and rings of the samples with a gradually changed indium deposition amount (a nominal amount of 1.35ML) obtained from AFM images as a function of indium deposition amount (b) variation of the second derivative RD spectra of the same series of samples, The results from up to down are for sample 1 to sample 16, respectively. The squares and triangles indicate the wavelengths of the LH related transition and GaAs band edge, respectively. [79]

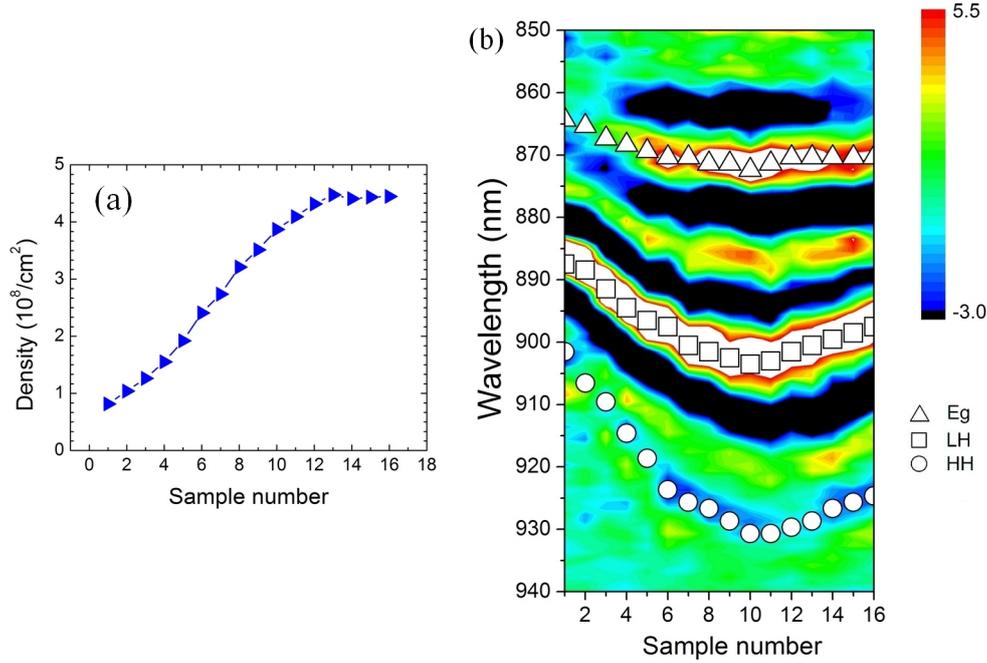

Figure 23: (a) Average densities of InAs rings from AFM measurement; (b) variation of the second derivative RD spectra with respect to wavelength with the samples indicated by the color contrast. The squares, circles, and triangles indicate the wavelengths of the LH related transition, HH related transition, and GaAs band edge, respectively. [78]

# 5 Summary

The recent progress of applying RDS technique in investigating the buried WLs has been reviewed in this article. The WL configurations can be extracted with the cross sectional imaging methods and various kinds of spectroscopy methods. RDS is more continent to achieve this goal. During the recent years, the evolution processes of WLs in different growth stages, the desorption of indium atoms and the effect of growth interruption in SK growth mode are systematically investigated with RDS. Besides, RDS can be also applied to study the buried two-dimensional structures grown with DE methods. Although some meaningful results of the profile and evolution of the WLs have been obtained with RDS and other techniques, an in-depth understanding of the segregation and some details of the substance distribution in the WLs are still in need for the controllable QD formation and physical properties.

# 6 Acknowledgements

The work was supported by the National Natural Science Foundation of China (No. 60625402 and 60990313), and the 973 program (2006CB604908 and 2006CB921607).